\documentstyle[12pt]{article}
\begin{document}

\catcode`@=11
\long\def\@caption#1[#2]#3{\par\addcontentsline{\csname
  ext@#1\endcsname}{#1}{\protect\numberline{\csname
  the#1\endcsname}{\ignorespaces #2}}\begingroup
    \small
    \@parboxrestore
    \@makecaption{\csname fnum@#1\endcsname}{\ignorespaces #3}\par
  \endgroup}
\catcode`@=12
\newcommand{\newc}{\newcommand}
\newc{\rem}[1]{{\bf #1}}
\newc{\gsim}{\lower.7ex\hbox{$\;\stackrel{\textstyle>}{\sim}\;$}}
\newc{\lsim}{\lower.7ex\hbox{$\;\stackrel{\textstyle<}{\sim}\;$}}
\newc{\gev}{\,{\rm GeV}}
\newc{\mev}{\,{\rm MeV}}
\newc{\ev}{\,{\rm eV}}
\newc{\kev}{\,{\rm keV}}
\newc{\tev}{\,{\rm TeV}}
\def\det{\mathop{\rm det}}
\def\tr{\mathop{\rm tr}}
\def\Tr{\mathop{\rm Tr}}
\def\Im{\mathop{\rm Im}}
\def\Re{\mathop{\rm Re}}
\def\bR{\mathop{\bf R}}
\def\bC{\mathop{\bf C}}

\newcommand{\Z}{Z}

\def\lie{\mathop{\hbox{\it\$}}} 
\newc{\sw}{s_W}
\newc{\cw}{c_W}
\newc{\swsq}{s^2_W}
\newc{\cwsq}{c^2_W}
\newc{\mgrav}{m_{3/2}}
\newc{\mz}{M_Z}
\newc{\mpl}{M_{pl}}
\def\ux{U(1)$_X$}
\def\beq{\begin{equation}}
\def\eeq{\end{equation}}
\def\bea{\begin{eqnarray}}
\def\eea{\end{eqnarray}}
\def\bi{\begin{itemize}}
\def\ei{\end{itemize}}
\def\benum{\begin{enumerate}}
\def\eenum{\end{enumerate}}

%
\def\boxeqn#1{\vcenter{\vbox{\hrule\hbox{\vrule\kern3pt\vbox{\kern3pt
\hbox{${\displaystyle #1}$}\kern3pt}\kern3pt\vrule}\hrule}}}
%

\def\qed#1#2{\vcenter{\hrule \hbox{\vrule height#2in
\kern#1in \vrule} \hrule}}
\def\half{{\textstyle{1\over2}}} 
\newc{\ie}{{\it i.e.}}          \newc{\etal}{{\it et al.}}
\newc{\eg}{{\it e.g.}}          \newc{\etc}{{\it etc.}}
\newc{\cf}{{\it c.f.}}
\def\CAG{{\cal A/\cal G}}
\def\CA{{\cal A}} \def\CB{{\cal B}} \def\CC{{\cal C}} \def\CD{{\cal D}}
\def\CE{{\cal E}} \def\CF{{\cal F}} \def\CG{{\cal G}} \def\CH{{\cal H}}
\def\CI{{\cal I}} \def\CJ{{\cal J}} \def\CK{{\cal K}} \def\CL{{\cal L}}
\def\CM{{\cal M}} \def\CN{{\cal N}} \def\CO{{\cal O}} \def\CP{{\cal P}}
\def\CQ{{\cal Q}} \def\CR{{\cal R}} \def\CS{{\cal S}} \def\CT{{\cal T}}
\def\CU{{\cal U}} \def\CV{{\cal V}} \def\CW{{\cal W}} \def\CX{{\cal X}}
\def\CY{{\cal Y}} \def\CZ{{\cal Z}}
\def\grad#1{\,\nabla\!_{{#1}}\,}
\def\gradgrad#1#2{\,\nabla\!_{{#1}}\nabla\!_{{#2}}\,}
\def\partder#1#2{{\partial #1\over\partial #2}}
\def\secder#1#2#3{{\partial^2 #1\over\partial #2 \partial #3}}
\def\ltap{\ \raise.3ex\hbox{$<$\kern-.75em\lower1ex\hbox{$\sim$}}\ }
\def\gtap{\ \raise.3ex\hbox{$>$\kern-.75em\lower1ex\hbox{$\sim$}}\ }
\def\gl{\ \raise.5ex\hbox{$>$}\kern-.8em\lower.5ex\hbox{$<$}\ }
\def\roughly#1{\raise.3ex\hbox{$#1$\kern-.75em\lower1ex\hbox{$\sim$}}}
\def\slash#1{\rlap{$#1$}/} 
\def\dsl{\,\raise.15ex\hbox{/}\mkern-13.5mu D} 
\def\delsl{\raise.15ex\hbox{/}\kern-.57em\partial}
\def\Ksl{\hbox{/\kern-.6000em\rm K}}
\def\Asl{\hbox{/\kern-.6500em \rm A}}
\def\Dsl{\hbox{/\kern-.6000em\rm D}} 
\def\Qsl{\hbox{/\kern-.6000em\rm Q}}
\def\gradsl{\hbox{/\kern-.6500em$\nabla$}}

%
\let\al=\alpha
\let\be=\beta
\let\ga=\gamma
\let\Ga=\Gamma
\let\de=\delta
\let\De=\Delta
\let\ep=\varepsilon
\let\ze=\zeta
\let\ka=\kappa
\let\la=\lambda
\let\La=\Lambda
\let\del=\nabla
\let\si=\sigma
\let\Si=\Sigma
\let\th=\theta
\let\Up=\Upsilon
\let\om=\omega
\let\Om=\Omega
\def\ph{\varphi}
\newdimen\pmboffset
\pmboffset 0.022em
\def\oldpmb#1{\setbox0=\hbox{#1}%
 \copy0\kern-\wd0
 \kern\pmboffset\raise 1.732\pmboffset\copy0\kern-\wd0
 \kern\pmboffset\box0}
\def\pmb#1{\mathchoice{\oldpmb{$\displaystyle#1$}}{\oldpmb{$\textstyle#1$}}
        {\oldpmb{$\scriptstyle#1$}}{\oldpmb{$\scriptscriptstyle#1$}}}
\def\bar#1{\overline{#1}}
\def\vev#1{\left\langle #1 \right\rangle}
\def\bra#1{\left\langle #1\right|}
\def\ket#1{\left| #1\right\rangle}
\def\abs#1{\left| #1\right|}
\def\vector#1{{\vec{#1}}}
\def\inv{^{\raise.15ex\hbox{${\scriptscriptstyle -}$}\kern-.05em 1}}
\def\pr#1{#1^\prime}  
\def\lbar{{\lower.35ex\hbox{$\mathchar'26$}\mkern-10mu\lambda}} 
\def\e#1{{\rm e}^{^{\textstyle#1}}}
\def\ee#1{\times 10^{#1} }
\def\imp{~\Rightarrow}
\def\coker{\mathop{\rm coker}}
\let\p=\partial
\let\<=\langle
\let\>=\rangle
\let\ad=\dagger
\let\txt=\textstyle
\let\h=\hbox
\let\+=\uparrow
\let\-=\downarrow
\def\dot{\!\cdot\!}
\def\vfilll{\vskip 0pt plus 1filll}
%

\begin{titlepage}
\begin{flushright}
{IASSNS--HEP--00/25\\ 
CERN--TH/2000--108\\
EFI--2000--17 \\ 
hep-th/0005276\\ }
\end{flushright}
\vskip 2cm
\begin{center}
{\Large Saltatory Relaxation of the Cosmological Constant
\footnote{{feng@ias.edu, jmr@nxth04.cern.ch,
sethi@theory.uchicago.edu, wilczek@ias.edu}}}

\vskip 0.3cm

{\bf Jonathan L. Feng$^*$, John March-Russell$^{\dagger}$, Savdeep
Sethi$^{\ast \, \ddag}$} 
\vskip 0.1in 
{\bf and Frank Wilczek$^*$} 
\vskip 0.5in 
$^*$ School of Natural Sciences, Institute for Advanced Study\\ 
Princeton, NJ 08540  USA 
\vskip 0.15in 
$^{\dagger}$ Theory Division, CERN, CH-1211, Geneva 23, Switzerland
\vskip 0.15in 
$^{\ddag}$ Enrico Fermi Institute, University of Chicago \\ 
Chicago, IL 60637  USA 

\end{center}

\vspace*{.2in}

\begin{abstract}
We modify and extend an earlier proposal by Brown and Teitelboim to
relax the effective cosmological term by nucleation of branes coupled
to a three-index gauge potential.  Microscopic considerations from
string/M theory suggest two major innovations in the framework.
First, the dependence of brane properties on the compactification of
extra dimensions may generate a very small quantized unit for jumps in
the effective cosmological term.  Second, internal degrees of freedom
for multiply coincident branes may enhance tunneling rates by
exponentially large density of states factors.  These new features
essentially alter the relaxation dynamics.  By requiring stability on
the scale of the lifetime of the universe, rather than absolute
stability, we derive a non-trivial relation between the supersymmetry
breaking scale and the value of the cosmological term.  It is
plausibly, though not certainly, satisfied in Nature.
\end{abstract}

\end{titlepage}

\section{Introduction}

Although both the Casimir effect of quantum theory and the existence
of symmetry-breaking condensates in both the strong and the
electroweak sectors of the Standard Model indicate that empty space is
a dynamical medium that `ought to' have a large mass density, gravity,
which couples universally to mass, does not reveal it.  This is the
problem of the cosmological term~\cite{weinberg}.  Various mechanisms
have been proposed to address this problem, but so far none has won
wide acceptance.  This situation is especially challenging for string
theory, and its conjectured non-perturbative definition M theory, in
as much as string theory is proposed as a fully specified dynamical
theory of gravity.

An interesting approach to the solution of the cosmological term
problem is the proposal that it is relaxed by jumps (saltation)
associated with some rather exotic dynamics.  There is an important
conceptual advantage to having the relaxation connected to some ---
necessarily exotic --- dynamics that responds only to a source taking
the form of an effective cosmological term.  For if the dynamics
responds to several influences, it is difficult to see how a
particularly simple value for one partial determinant of its behavior
can become overwhelmingly preferred.  A model for this logic is the
advantage one gains from Peccei-Quinn symmetry in relieving the
$\theta$-problem of QCD.  Indeed, that symmetry produces an exotic
dynamics (the axion field) which responds only to a source having the
form of an effective $\theta$ term.  Note that this logic also
applies, in connection with {\em continuous\/} relaxation of scalar
fields, to self-interactions (including kinetic terms), as has been
quantified by Weinberg~\cite{weinberg}.

Saltation in the effective cosmological term has been considered in
the context of stepwise false vacuum decay~\cite{CC,CdL} in a
quasi-periodic staircase potential by Abbott~\cite{abbott}, or
alternatively by Brown and Teitelboim (BT) through nucleation of
fundamental membrane degrees of freedom~\cite{BT}.  Since the membrane
formulation is close in spirit to the ones that appear to arise
naturally in string theory, we shall phrase our discussion in its
framework.

The essential ingredients in the BT model are a fundamental membrane
degree of freedom (in modern language, a 2-brane) and a 4-form gauge
field strength $F_4$ (deriving from a 3-form potential).  The
world-volumes of the membranes couple to the 3-form potential.  The
field strength $F_4$ has no other couplings and no local dynamics in
the four-dimensional spacetime, but its expectation value contributes
to the effective cosmological term.  The presence of an expectation
value for $F_4$ induces the nucleation of the 2-brane to which it
couples, in a manner analogous to the Schwinger mechanism for electric
field decay through nucleation of $e^+ e^-$ pairs.  When a membrane is
nucleated, say as a spherical shell, the effective value of the
cosmological term on the inside differs from its previous value (the
value on the outside) by an amount proportional to the coupling
constant of the membrane.  If a membrane of the correct sign is
nucleated, the contribution of the 4-form to the effective value of
the cosmological term will be reduced.  For a suitable value of
$\vev{F_4}$, this contribution can cancel that arising from the
combined effects of all other degrees of freedom in the theory.  If
the steps between adjacent false vacua are sufficiently small, and if
there is a reason why vacuum decay stops or slows down dramatically as
$\La_{\rm eff} \rightarrow 0$, this mechanism could in principle relax
a large microscopic cosmological term (arising from all sources other
than the membrane-$F$ field dynamics) to a value within observational
bounds.

In their pioneering work, Abbott, and Brown and Teitelboim, postulated
dynamical entities {\em ad hoc}, with no richer context to enhance
their credibility or connect them to other problems and facts of
physics.  On the other hand, developments in string theory in the past
few years have emphasized the importance of extended objects ---
branes --- of various types.  It therefore seems appropriate to
consider whether these branes might allow improved mechanisms for
saltatory relaxation of the cosmological term.  We believe that this
is the case, for several reasons:

\bi
\item
A wide variety of membranes play a fundamental role in string
theory. There are corresponding gauge fields, which naturally couple
(only) to these extended objects.
\item
Since the theory is naturally formulated in higher dimensions, the
couplings of these membranes as seen in four dimensions are not fixed
and quantized, but rather are determined in terms of the fundamental
(fixed, quantized) couplings together with properties of the
extra-dimensional compactification.
\item
Similarly, the effective tension as seen in four dimensions depends on
the properties of the extra-dimensional compactification.
\item
There are significant and, in suitable cases, exponentially large,
density of states factors associated with semi-classical brane
processes.
\item
Small tension, which may be favored for dynamical reasons, and large
density of states factors make possible rapid relaxation of the
cosmological constant.  \ei

We should stress that regardless of whether membrane nucleation is the
final solution, or perhaps an ingredient, in solving the riddle of the
cosmological constant, nucleation processes involving extended objects
generically occur in string theory.  Studying these processes is
likely to give some insight into the question of vacuum selection and
into the question of a background independent formulation of string
theory.

In the following section, we review the BT formalism for relaxation of
the cosmological constant through brane nucleation and the primary
obstacles encountered.  In the next two sections, we discuss features
of string theory that may alleviate these difficulties: in
Sec.~\ref{sec:charge} we describe the possibility of small charge
densities and tensions arising from compactification, and in
Sec.~\ref{sec:degeneracy} we note the relevance of exponentially large
density of states factors.  Motivated by these features, we describe
two possible scenarios for the cosmological constant in
Sec.~\ref{sec:scenarios}.  In Sec.~\ref{sec:discussion}, we discuss a
number of outstanding issues and summarize.

\section{Relaxation Dynamics}
\label{sec:BT}

\subsection{Basic mechanism}

We now recall the basic dynamics of the BT mechanism~\cite{BT}, which
we have been able to express in a somewhat simplified fashion. For
other recent discussions of membrane nucleation, see~\cite{others}.

Consider gravity in $D=4$ spacetime dimensions with a 2-brane
$X^{\alpha}$ coupled to a 3-form gauge potential $A_3$.  The Minkowski
action is
\begin{eqnarray}
S_M &=& - \tau_2 \int d^3 \xi \sqrt{-\det g_{ab}}
+ \frac{\rho_2}{6} \int d^3 \xi A_{\alpha \beta \gamma}
 \frac{\partial X^{\alpha}}{\partial \xi^a}
 \frac{\partial X^{\beta}}{\partial \xi^b}
 \frac{\partial X^{\gamma}}{\partial \xi^c}
\varepsilon^{abc} \nonumber \\ && - \frac{1}{48} \int d^4 x
\sqrt{-g} F_{\alpha\beta\gamma\delta} F^{\alpha\beta\gamma\delta}
+ \frac{1}{6} \int d^4 x \partial_\alpha \left[ \sqrt{-g}
F^{\alpha \beta \gamma \delta} A_{\beta \gamma \delta} \right]
\nonumber \\ && + \frac{1}{2} \int d^4 x \sqrt{-g} M^2 (R - 2
\la) - M^2 \oint d^3 x \sqrt{h} K \ , \label{Sm}
\end{eqnarray}
where the $\xi^a$ parameterize the membrane world-volume, and $g_{ab}
= \partial_a X^{\alpha} \partial_b X_{\alpha}$ is the induced
world-volume metric.  The surface integral is over spacetime
boundaries with $h$ and $K$ the induced metric and extrinsic
curvature, respectively.  This term and the total derivative integral
ensure that the action has well-defined functional derivatives with
respect to the metric and gauge field.  An important point is that in
four dimensions the 4-form field strength contains no independent
propagating degrees of freedom, its value, up to a constant, being
fully determined by the background of sources charged with respect to
$A_{\alpha \beta \gamma}$.

The parameters entering this action, and their mass dimensions in
$D=4$, are
\beq
\begin{array}{lcl}
\mbox{2-brane tension} & \tau_2 & 3 \\
\mbox{2-brane charge density} & \rho_2 & 2 \\
\mbox{(Reduced) Planck mass} & M  & 1 \\ \mbox{Bare cosmological
constant} & \qquad \la \qquad & 2 \ \ . \\
\end{array}
\eeq
Numerically, $M = (8\pi G)^{-1/2} = 2.4 \times 10^{18}\gev$. Here, in
agreement with BT, we use the canonical --- positive energy --- sign
for the $F^2$ term. In Sec. 3, we shall see that this is appropriate
for the branes relevant to us.

Rotating to Euclidean space, we find
\begin{eqnarray}
S_E &=& \tau_2 \int d^3 \xi \sqrt{\det g_{ab}}
+ \frac{\rho_2}{6} \int d^3 \xi A_{\alpha \beta \gamma}
 \frac{\partial X^{\alpha}}{\partial \xi^a}
 \frac{\partial X^{\beta}}{\partial \xi^b}
 \frac{\partial X^{\gamma}}{\partial \xi^c}
\varepsilon^{abc} \nonumber \\
&& - \frac{1}{48} \int d^4 x \sqrt{g} F_{\alpha\beta\gamma\delta}
F^{\alpha\beta\gamma\delta}
+ \frac{1}{6} \int d^4 x \partial_\alpha \left[ \sqrt{g} F^{\alpha
\beta \gamma \delta} A_{\beta \gamma \delta} \right] \nonumber \\
&& + \int d^4 x \sqrt{g} \frac{1}{2} M^2 (-R + 2 \la)
+ M^2 \oint d^3 x \sqrt{h} K \ .
\label{SE}
\end{eqnarray}
The sign of the $F^2$ term in Eq.~(\ref{SE}) depends on the
Euclideanization procedure.  Here, following Ref.~\cite{BT}, we make
the conventional rotations $x^0 \rightarrow -i x^0$, $X^0 \rightarrow
-i X^0$ for time-like quantities, but take $A^{0 \mu_2 \ldots
\mu_{D-1}} \to A^{0 \mu_2 \ldots \mu_{D-1}}$ and $A^{\mu_1 \ldots
\mu_{D-1}} \to i A^{\mu_1 \ldots \mu_{D-1}}$ for the space-like
components so that the field strength $F$ is invariant. Alternatively,
one may adopt the prescription $A^{0 \mu_2 \ldots \mu_{D-1}} \to -i
A^{0 \mu_2 \ldots \mu_{D-1}}$ and keep the space-like components
invariant.  In this case, the sign of the $F^2$ term in Eq.~(\ref{SE})
changes.  However, the field strength is not invariant under this
prescription; taking $\langle F_4 \rangle$ in Eq.~(\ref{SE}) to be
pure imaginary~\cite{HT} leaves the following analysis unchanged.

The instanton solution is a membrane that divides space into two
regions, an outside $O$ and an inside $I$.  In each region, the field
strength is a constant
\beq
\vev{F_{O,I}^{\alpha\beta\gamma\delta}} = \frac{c_{O,I}}{\sqrt{g}}
\varepsilon^{\alpha\beta\gamma\delta} \ ,
\eeq
and the field strengths are matched across the membrane boundary via
\beq
c_I = c_O - \rho_2 \ .
\eeq
The effective cosmological terms are
\beq
\La_{O,I} = \la + \frac{1}{2 M^2} c_{O,I}^2 \ ,
\label{cceff}
\eeq
where the field strength contribution follows from Einstein's
equations.  Alternatively, it may be `read off' from the action if one
is careful to include the on-shell contribution from the total
derivative term, which is double in magnitude and opposite in sign
relative to the usual $F^2$ term.  {}From Eq.~(\ref{cceff}), it is
clear that if the bare cosmological term is to be canceled, it must be
negative, and we therefore assume $\la<0$.

The tunneling probability is $P \sim e^{-B}$, where the bounce action
\cite{CC,CdL} for this false vacuum decay is
\beq
B = \left\{ \begin{array}{ll}
\infty\ , & \rm{if}~ \La_O, \alpha_O < 0 \\
12 \pi^2 M^2 \left[ \frac{1}{\La_O} (1-b \alpha_O) -
\frac{1}{\La_I} (1-b \alpha_I) \right] \ , & \rm{otherwise} \ .
\end{array} \right.
\label{bounce}
\eeq
Here the bubble radius, defined so that the area of the bubble slice
when continued back to Minkowski signature is $4\pi b^2$,  is
\beq
b = \frac{1}{\sqrt{\frac{1}{3} \La_{O} + \alpha_{O}^2 }}=
\frac{1}{\sqrt{\frac{1}{3} \La_{I} + \alpha_{I}^2 }} \ ,
\eeq
and
\begin{eqnarray}
\alpha_{O \atop I} &=& \frac{1}{3xM} \left[ c_O - \left(\frac{1}{2}
\pm \frac{3}{4} x^2 \right) \rho_2 \right] \\
\La_I &=& \La_O + \frac{1}{2M^2} ( \rho_2^2 - 2 \rho_2 c_O) \\
x &=& \frac{\tau_2}{\rho_2 M} \ .
\label{relations}
\end{eqnarray}

In the following sections, we will often work in Planck units with
$M\equiv 1$.

\subsection{Naturalness}
\label{sec:naturalness}

The value of the cosmological term at present is
\beq
\La_{\rm obs} M^2 \lsim (2\times 10^{-3}\ev)^4 \sim 10^{-120}
M^4 \ .
\eeq
Present data prefer a non-zero, positive value; for recent reviews,
see~\cite{reviews}. One virtue of saltatory relaxation is that, in
contrast to mechanisms involving symmetries or continuous relaxation,
a small non-zero value may emerge naturally as a consequence of a
non-vanishing jump size.

For the observed value to be natural in the framework of brane
nucleation, the spacing between allowed values of the effective
cosmological term, near the observed value, cannot be much larger.
This translates into the condition
\beq
\rho_2 \lsim \frac{\La_{\rm obs}}{|\la|^{1/2}} \ .
\label{fineness}
\eeq 
This is an extremely stringent condition on the microphysics, even for
plausible $|\la| \ll 1$, since the observed cosmological constant is
so small.  For example, even if the bare cosmological constant is
generated only at the TeV scale through low-energy supersymmetry
breaking so that $|\la| \sim 10^{-60}$, one still requires $\rho_2
\lsim 10^{-90}$, which translates into an associated mass scale of
$10^{-18}\ev$.  The requirement of such a small coupling is, at best,
unsettling, and one might hope for an explanation in some more
fundamental framework.  This is especially true in string theory,
where, in the absence of such an explanation, one naively expects
couplings to be of order the Planck scale.

As an aside, note that throughout this study, we consider evolution of
the cosmological term down the staircase of values allowed by brane
nucleation with fixed charge density $\rho_2$, and thus with
essentially fixed step size.  However, very small fractional changes
in $\rho_2$, of order
\beq
{\delta \rho_2 \over \rho_2} \lsim {\La_{\rm obs} \over |\la|}
\ ,
\eeq
assuming $\rho_2 \ll \sqrt {|\la|}$, are sufficient to bring the
observed value of the effective cosmological term into range by
distorting the size of the steps near $\La=0$.  Another possibility is
that the entire staircase moves up or down by some suitably small
amount.  Although we will not attempt to exploit these features here,
they may play an important role in future work, since, as we shall
emphasize below, $\rho_2$ is in principle a dynamical variable.  It
depends, in particular, on the expectation values of the string theory
compactification moduli, including the dilaton.

\subsection{Absolute Stability}
\label{sec:stability}

The cosmological term must not only relax to within its observational
bounds, but it must also stop evolving once it reaches this interval.
For de~Sitter space, additional bubble nucleations are always
possible. However, for $\La_O < 0$, transitions can take place only
when $\alpha_O > 0$.  It is not hard to show that this constraint,
along with the condition $\frac{1}{3} \La_O + \alpha_O^2 >0$, implies
that further lowering of the effective cosmological term will not
occur beyond the first anti-de~Sitter step if the tension is
sufficiently large.  This remarkable result is closely related to the
Coleman-de~Luccia~\cite{CdL} gravitational suppression of false vacuum
decay.  BT hypothesized that our universe is at the endpoint of such
an evolution.

A sufficient condition to ensure absolute vacuum stability is
$\tau_2^2 > \frac{4}{3} \rho_2 c_O$ or, in terms of the tension to
charge density ratio,

\beq
x = \frac{\tau_2}{\rho_2} \gsim \sqrt{\frac{|\la|}{\La_{\rm
obs}}} \ .
\eeq
Thus, even for the smallest plausible $|\la|$, a large hierarchy
between tension and charge density is required.

More problematic still, BT showed that, upon combining the stability
and naturalness conditions, the time required to reach the endpoint is
excessively large, so large that even the very slow inflation that
occurs in the penultimate vacuum would leave the universe entirely
devoid of matter and energy.

\section{Saltation in String Theory: Tension and Charge Density}
\label{sec:charge}

\subsection{Framework}

In its long wavelength approximation, M theory supports BPS M2-branes
and M5-branes~\cite{townsend,Mlectures,JPbook}.  The M2-branes couple
electrically to the 3-index gauge potential of $11$-dimensional
supergravity, which we denote by $A_3$. We define a dual gauge
potential $A_6$ in the standard way:
\bea
F  = dA_3 =  * F_7 =  * dA_6 \ .
\label{dualpotential}
\eea
The M5-branes couple magnetically to $A_3$, or directly to $A_6$.  In
terms of the 11-dimensional Planck scale $l_p$, M2-branes have a
tension $T_2 \sim (l_p)^{-3}$ while M5-branes have a tension $T_5 \sim
(l_p)^{-6}$.

The simplest way to arrive at a $4$-dimensional world is via
compactification on a $7$-dimensional internal space ${\cal{M}}$.  We
obtain 2-branes from either the fundamental M2-branes of M theory, or
by wrapping M5-branes on a 3-cycle $a_3$ of the internal space
${\cal{M}}$.  Our spectrum of 4-forms in spacetime comes about in the
following way: let $\omega^i$ be a basis for $H^3(\CM, \Z)$. We
require integer forms so that the resulting 4-form fields satisfy
Dirac quantization.  We can then expand $F_7$ in this basis,
\beq
F_7 = \widetilde{F}_i \wedge \omega^i \ .
\eeq
Note that we get many 4-forms from a generic compactification of this
kind.  There is a natural geometric picture associated to this
expansion. To each form $\omega^i$, we can associate a dual $3$-cycle
on which we wrap an M5-brane to obtain a 2-brane.

After setting the fermions to zero, the 11-dimensional metric
satisfies the equation of motion
\beq
R_{ab} - \frac{1}{2} g_{ab} R  + \frac{1}{6} \left( F_{acde}
F_{b}^{\, cde}  - \frac{1}{8} g_{ab} F_{cdef} F^{cdef} \right)=0 \ .
\label{EOM}
\eeq 
The sign of the four-form contribution is fixed by 11-dimensional
supergravity.  For both kinds of effective 2-brane described above,
the contribution of $F$ (or $\widetilde{F}$) to the effective
cosmological constant is positive. We therefore require a negative
bare cosmological constant $\lambda$, as assumed in section 2.

If $\CM$ is circle-fibered then we can reduce M theory on the fiber to
obtain various string compactifications. For example, if we take $\CM$
to have the form $S^1 \times K$ then for a small $S^1$ of size
$R_{11}$, M theory reduces to type IIA string theory on
$K$~\cite{townsend}.  The string scale $\alpha'$ and string coupling
$g_s$ are given in terms of $l_p$ and $R_{11}$,
\beq
\alpha' = l_p^3 /R_{11}, \quad g_s = (R_{11}/l_p)^{3/2}.
\eeq

Type IIA string theory contains an NS-NS (Neveu-Schwarz) 3-form field
strength $H_3$ which measures fundamental string charge. The magnetic
dual $H_7 = *H_3$ measures the charge of NS 5-branes. These 5-branes
have a tension proportional $1/g_s^2$ and so are very heavy at weak
string coupling. In addition, there are Ramond-Ramond (RR) field
strengths $F_2$ and $F_4$ (and a non-dynamical $F_{10}$~\cite{polRR})
together with their 8-form and 6-form (and 0-form) magnetic duals.
These $(p+2)$-form RR field strengths couple electrically to dynamical
Dirichlet $p$-branes~\cite{polRR,Dbranes} through the world-volume
action
\begin{equation}
\int d^{p+1}\xi \, A_{p+1}\ .
\end{equation}
In type IIA, we therefore have the additional possibility of wrapped
D6-branes and D8-branes giving rise to effective 2-branes.  In (9+1)
dimensions, the low-energy string and D$p$-brane effective action is
\bea
S_{10} &=& \frac{1}{2\ka_{10}^2} \int d^{10}x  \sqrt{-G}
\left[ e^{-2\phi} \left( R + 4(\nabla\phi)^2 \right)
- \sum {1\over 2n!} F_n^2 +\ldots \right] \nonumber \\
&& - T_p \int d^{p+1}\xi\, e^{-\phi} \sqrt{-\det G_{ab}} +
\rho_p\int d^{p+1}\xi\, A_{p+1} +\ldots \ ,
\label{Dpstringframe}
\eea
where $\phi$ is the dilaton, $G_{ab}$ is the induced metric on the
brane, and only the relevant bosonic terms are displayed.  The tension
and charge density of type II D$p$-branes is
\beq
T_p^2 = \rho_p^2 = \frac{\pi}{\ka_{10}^2} (4\pi^2 \al')^{3-p} \ .
\eeq
This may be written more conveniently by using the relation
$\ka_{10}^2 = 2^6 \pi^7 \al'^4$ and defining the string length
$\ell_s$ in terms of the fundamental string tension through $T_{\rm
F1} = 1/(2\pi\alpha') = 2\pi/\ell_s^2$.  With these conventions,
\beq
T_p = \rho_p = \frac{2\pi}{\ell_s^{p+1}} \ .
\eeq
(Note that there is no factor of $1/g_s$ in these expressions because
of the conventions employed in Eq.~(\ref{Dpstringframe}).)

The 4-dimensional physical brane tensions and charge densities are
proportional to $T_p$ and $\rho_p$.  As we discuss in the next
subsections, the exact relations depend on the complicated dynamics of
compactification which are not well understood.  For this reason, when
we consider scenarios in Sec.~\ref{sec:scenarios}, we will take a
4-dimensional effective action approach, and simply assume suitable
values for the 2-brane tension and charge density.  In the rest of
this section, however, we explore the possible effects of
compactification on these quantities.

There are also other ways of arriving at a 4-dimensional world that do
not fit into the above framework. For example, compactifying F
theory~\cite{vafa} on a Calabi-Yau 4-fold~\cite{sethione} naturally
gives a large class of N=1 compactifications. The 4-fold must be
elliptically-fibered with a section. If $B$ denotes the 6-dimensional
base of the fibration then over each point of $B$, the structure of
the elliptic fibration specifies a torus.  By F theory compactified on
the 4-fold, we mean type IIB string theory compactified on
$B$. However, the complexified string coupling constant $\tau$ is not
constant but varies over $B$ in a manner determined by the shape of
the torus fiber. Therefore, the string coupling constant is typically
not a tunable modulus for these compactifications.

To determine, the spectrum of 2-branes for an F theory
compactification, it is natural to use the duality between M theory on
$T^2$ and the type IIB string. From this duality, we see that 2-branes
arise only from M5-branes. An M5-brane wrapping the elliptic fiber and
a 1-cycle in $B$ would look like a D3-brane wrapping the 1-cycle in
$B$. There are additional possibilities. For example, an M5-brane with
a leg on the elliptic fiber and 3 legs on $B$ can give rise to a
2-brane. From the IIB perspective, it would appear to be a combination
of $(p,q)$ 5-branes wrapping a 3-cycle of $B$. In backgrounds such as
these with N=1 supersymmetry, we also need to worry about the
spacetime superpotential.  However, our subsequent discussion will be
largely classical. Our aim is to be as simple as possible in order to
isolate the aspects of compactifications that are most relevant for
the brane nucleation mechanism.
 
We also note that there is a much broader class of compactifications
that involve background fluxes. These compactifications, which are
actually generic string compactifications, typically warp
spacetime~\cite{beckers,sethitwo,greene}.  These backgrounds can
potentially give rise to the kind of novel infra-red physics that
could make saltatory relaxation a viable mechanism for reducing the
cosmological constant.

\subsection{Direct descent}

One might suppose that M2-branes are made to order for saltatory
relaxation of the cosmological term. However, as we shall see, they do
not have the right properties, at least when taken in their
straightforward form.  The wrapped M5-branes appear more promising.

Let us first consider the case of a D2-brane in 10 dimensions that
descends directly to a 2-brane in 4 dimensions.  Our discussion will
be in the context of 10-dimensional type IIA supergravity compactified
on a Calabi-Yau manifold $K$ with string-frame volume $V_6$.  This
compactification preserves N=2 supersymmetry.  The physical effective
tensions and charge densities are then determined in 4-dimensional
Einstein frame, where the gravitational action takes the conventional
Einstein-Hilbert form.  This frame follows after performing the Weyl
rescaling $g_{\mu\nu} \to e^{2\phi} g_{\mu\nu}$ on the 4-dimensional
metric $g_{\mu\nu}$.  The 4-dimensional action is then
\bea
S_4 &=& \frac{V_6}{2\ka_{10}^2} \int d^4 x  \sqrt{-g}
\left[ R - 2 (\nabla\phi)^2 - \sum {1\over 2n!} e^{-2 (n-2) \phi}
F_n^2 + \ldots \right] \nonumber \\
&& - T_p \int d^{p+1}\xi\, e^{p\phi} \sqrt{-\det g_{ab}} +
\rho_p\int d^{p+1}\xi\, A_{p+1} +\ldots \ .
\eea
The 4-dimensional effective tension is $\tau_p \bigr|_{\rm 4D, eff} =
T_p g_s^p$, where $g_s = e^{ \langle \phi \rangle}$ is the string
coupling.  To determine the effective charge density, note that the
$F_n^2$ kinetic terms are not canonically normalized.  Upon restoring
the normalization, we find a 4-dimensional effective charge density of
$\rho_p \bigr|_{\rm 4D, eff} = \sqrt{2} \rho_p g_s^p / M$, where the
reduced Planck mass is fixed by $M^2 = V_6 / \ka_{10}^2$.  The tension
to charge density ratio, in Planck units, is then $x = 1/\sqrt{2}$, as
we expect for BPS branes.

For the 2-brane case of interest,
\beq
\rho_2 \bigr|_{\rm 4D, eff} =
\frac{2\sqrt{2} \pi g_s^2}{M\ell_s^3} \ .
\label{rho2Eframe}
\eeq 
Clearly, to obtain a sufficiently small charge density, we must have
extreme values for $\ell_s$ and/or $g_s$.  For example, for the
canonical choice of string scale $\ell_s\simeq (10^{17}\gev)^{-1}$, we
find that a charge density of $\rho_2 \lsim 10^{-90}$ requires
$g_s\lsim 10^{-44}$.  Alternatively we could take $\ell_s$ to be a
larger length scale.  These non-canonical cases include the `large
extra dimension' scenario~\cite{AAHDD} with $g_s\sim 1$ and some
number of sub-millimeter dimensions.  However, given the success of
quantum field theory at colliders such as LEP and the Tevatron,
$\ell_s \lsim (\tev)^{-1}$ at the very best.  Although an improvement,
this still requires a tiny string coupling $g_s\lsim 10^{-23}$ to
generate a sufficiently small $\rho_2$.  It is possible that this
string coupling is unrelated to the gauge couplings of the Standard
Model; for example, the Standard Model may arise from some
non-perturbative sector of string theory.  However, if $g_s$ is
related to the Standard Model gauge couplings in a straightforward
way, it is difficult to understand how to accommodate the Standard
Model in such an extreme corner of string theory moduli space, much
less gauge coupling unification with $\al_{\rm unif}\simeq
1/25$~\cite{gauge}.

\subsection{Degenerating cycles}
\label{sec:degenerating}

A more promising alternative to direct descent branes are branes
wrapped on homology cycles.  These branes become tensionless when the
volume of these cycles approaches zero classically, as for
conifolds~\cite{conifold}. For compactifications with $N \geq 2$
supersymmetry, these branes give rise to BPS states and so quantum
corrections do not change this conclusion.

Specifically, if a $p$-brane of tension $T_p$ wraps a $k$-cycle $a_k$
of the compactification manifold, where $k \le p$ and the volume of
$a_k$ is ${\rm Vol}(a_k)$, then the result in the effective
4-dimensional theory is a $(p-k)$-brane of tension $\tau_{(p-k)} \sim
T_p \cdot {\rm Vol}(a_k)$.  In particular, for an effective 2-brane in
4 dimensions coming from the wrapping of a D$p$-brane on a cycle
$a_{p-2}$, the Einstein frame effective tension is
\beq
\tau_2\bigr|_{\rm 4D, eff}= T_p {\rm Vol}(a_{p-2}) g_s^2\ .
\label{densities}
\eeq
If ${\rm Vol}(a_{p-2})$ approaches zero, \ie, $a_{p-2}$ is a
degenerating cycle, a nearly tensionless object exists in the
4-dimensional theory.  The analogous formula for the tension of a
wrapped NS 5-brane just differs by a factor of $g_s$. This is
consistent with the higher-dimensional quantization rules for brane
properties.

What determines the charge density? As we saw in the case of the
direct reduction of an M2-brane to a D2-brane, the kinetic terms for
the $3$-form gauge-field become moduli-dependent. The term,
$$\int d^{p+1}\xi \, A_{p+1}\ ,$$ is reduced on integer classes and so
contains no moduli dependence. For purposes of determining the scaling
of the charge density with the moduli, we only need the behavior of
the moduli space metric near the singularity. At the level of
classical geometry, this is determined by the behavior of the
Weil-Petersson metric.  For N=2 compactifications, this metric
behavior is not changed by quantum corrections, although it is
certainly renormalized in situations with less supersymmetry.  Let us
take the case of the conifold where an $S^3$ shrinks to zero size in
$K$.  We choose coordinates for the moduli space so that the
singularity is located at $Z=0$. The tension of a wrapped brane goes
like,
$$ \tau_2\bigr|_{\rm 4D, eff} \sim |Z|\ . $$ 
However, the moduli space metric scales like $\ln (Z)$ and so the
charge density behaves like~\cite{concandelas},
$$  \rho_2 \bigr|_{\rm 4D, eff} \sim 1/\sqrt{| \ln (Z)|}\ .$$
Thus although the tensions can be vanishingly small, the charge
densities generated are only slightly smaller than in the direct
descent case.  We would therefore need to stabilize $Z$ at extremely
small values in order to obtain an acceptably small charge
density.\footnote{We thank J.~Polchinski for correcting an error in an
earlier version of this paper, and J.~Maldacena for discussions.}

The log-singularity in the metric has a simple space-time
interpretation as arising from integrating out a massless
hypermultiplet~\cite{conifold}. We desire a singularity at finite
distance in the moduli space whose behavior is worse than that of a
conifold.  It is not hard to see that in the context of N=2
compactifications, such singularities cannot arise by simply tuning
vector multiplet moduli. For example, one can try generalizations of
the conifold with multiple simple nodes.  These singularities are all
at finite distance in the moduli space.  However, in each case and in
general, it appears that gravity can be decoupled. The resulting
theory of vector multiplets has a moduli space metric with
singularities than metrically cannot be worse than logarithmic by
standard non-renormalization theorems.

Too little is currently known about the hypermultiplet moduli space to
decide whether a singularity with the right metric behavior
exists. Some results on classifying singularities at finite distance
have appeared in~\cite{hayakawa}.  However, when we relax the
condition of N=2 supersymmetry and consider the far larger class of
N=1 compactifications, it seems far more likely that a sufficiently
bad singularity exists (classically).  This is quite exciting since it
connects the possibility of novel infra-red physics with a mechanism
for reducing the cosmological constant. A recent F theory example with
unusual infra-red physics arising from a bad singularity appeared
in~\cite{morrison}. This kind of example certainly has the right
qualitative features, and it actually seems quite hard to rule out the
existence of a compactification with the features we desire. Clearly
more work along these lines is needed.

We have yet to discuss supersymmetry breaking.  One natural mechanism
worth mentioning in this context breaks supersymmetry by turning on
$RR$ and $NS$ fluxes on $K$. For example, we can give (3+1)-Lorentz
invariant expectation values to the RR fields $F_n$ by taking,
$$ \vev{F_2} =v^i_2\, \alpha^{(2)}_i, \qquad 
\vev{F_4} = v^i_4\, \alpha^{(4)}_i +
v_4\, \ep^{(4)}. $$
The $\alpha_i$ are harmonic forms on $K$, and $\ep^{(4)}$ is the
spacetime volume element. If we take the 10-dimensional Poincar\'e
duals of these expectation values $v^i_{(4,2)}$, we then find
expectation values of the general form $F = v^i_{(4,2)}
\ep^{(4)}\wedge {\widetilde \om}^{(2,4)}$, which couple magnetically
to the D$p$-branes.  Let us denote the expectation values collectively
by $v_a$.

Generically, when $v_a\neq 0$, supersymmetry is
broken~\cite{candelas,PS,michelson,taylorvafa,mayr00}, and a scalar
potential is generated that depends on the Calabi-Yau moduli $t_\al$
and the expectation values $v_a$.  What is interesting to us, however,
is that the potential naturally tends to drive us to singular
compactifications.  In some examples involving just RR fluxes, two
types of critical point for the scalar potential have been
found~\cite{PS,michelson}: either the Calabi-Yau runs away to infinite
volume, or the theory is driven to conifold-like points where homology
cycles of $K$ degenerate (classically approach zero volume).  The
minima tend to either break supersymmetry completely or restore the
full N=2.  Moreover, since the configurations with $v_a \neq 0$ are
not iso-potential, there are dynamical processes whereby the values of
the $v_a$'s change.  The $v_a$ `discharge' by the nucleation and
expansion of charged membranes, the D-branes of string theory.

Classically, the vacuum expectation values $v_a$ of the RR fields can
vary continuously.  Quantum mechanically, they satisfy quantization
rules that follow from the quantization of brane properties, which are
standard results in string theory~\cite{Mquant}.  These brane
properties are fixed, given information about the compactification and
the parameters of our low-energy effective Lagrangian.

There is an additional issue concerning this means of supersymmetry
breaking that deserves mention.  In the context of M theory on
$K\times S^1$, a class of 2-branes arise from M5-branes wrapping
3-cycles.  In type IIA, these 2-branes arise from either D4-branes
wrapping 2-cycles of the compactification space $K$, or from NS
5-branes wrapping 3-cycles of $K$. The two cases are quite
different. In the case of the wrapped D4-branes, we must also consider
D2-branes which can also wrap the shrinking 2-cycle. These wrapped
branes give rise to particle states with mass $\tau_0 \sim {\rm
Vol}(a_2)/\ell_s^3 \to 0$. With multiple collapsed nodes, we will find
an interacting gauge theory in $3+1$-dimensions. In this case, it
seems possible that the wrapped D4-branes will correspond to
collective excitations of the gauge theory. This is not an issue,
however, for the wrapped NS 5-branes since there are no BPS D3-branes
in type IIA string theory. These wrapped NS-branes give rise to
inherently non-perturbative stringy excitations.

The phenomenon of classically vanishing tensions arising from branes
wrapped on degenerating cycles is intriguing.  It is also worth noting
that the classical phenomenon of true degeneration and corresponding
tensionless, or massless, states is typically not realized in the full
quantum theory.  Instead the effective (3+1)-dimensional 2-brane will
have a dynamically generated non-perturbative tension, which may be
{\em exponentially small}.  It is known that in some cases a tension
is generated from the dynamics of the would-be massless particle
states arising from a D2-brane wrapping the 2-cycle.  These particle
states realize a non-Abelian gauge theory, presumably in a sector
hidden with respect to the Standard Model, whose low-energy
non-perturbative dynamics can break supersymmetry. (This sector is
conceptually distinct from hidden sectors postulated to provide
supersymmetry breaking for the supersymmetric Standard Model.)  This
leads to a potential for the volume modulus of the cycle, which
stabilizes it at a scale $\La \sim \exp(-8\pi^2/b_1
g_{YM}^2)M$~\cite{mayr96,mayr00,taylorvafa,mayrprivate}.  Once the
cycle is stabilized at this small scale, membranes wrapping this cycle
have a tension that is also proportional to $\La$ and thus can be very
small, even for $g_{YM} \sim 1$.  The precise conditions under which
very small tensions occur deserve much further study.

Some years ago, one of us made the numerical joke~\cite{joke}
\beq
\La_{\rm obs} M^ 2 \sim (e^{- \pi/\alpha_{\rm unif}} M)^4 \ ,
\eeq
with $\alpha_{\rm unif} \simeq 1/25$.  Our present considerations
suggest an intimate connection between non-perturbative effects and
the value of the observed cosmological term, which could conceivably
lead to a relation of just this form.

\section{Saltation in String Theory: Density of States}
\label{sec:degeneracy}

\subsection{Multi-bounce properties}

An essentially new feature is introduced by multi-bounce solutions
arising from coincident branes.  Such coincident branes support
low-energy internal degrees of freedom.  Thus there are potentially
large density of states factors accompanying their nucleation.
Calculations~\cite{BK,kiritsis} performed in the context of checking
duality between type I and heterotic SO(32) string theory demonstrate
that D-branes do make contributions that can be interpreted
semi-classically as incorporating degeneracy factors reflecting the
non-Abelian structure of coincident D-branes.  Another aspect of this
is that many coincident branes with large total charge can be
described in appropriate limits as `black' objects, similar to black
holes, with event horizons, and with associated Bekenstein-Hawking
(BH) entropy~\cite{KT}.

Consider first the case of $k$ coincident D3-branes.  Such a
configuration possesses a U($k$) super Yang-Mills (SYM) gauge theory
on its world-volume.  In the limit where the interactions with the
bulk string theory are weak, and where the temperature (or excitation
energy) of the SYM is small, one can compute the entropy of this
system.

When the effective SYM gauge coupling $g^2_{\rm eff}\simeq k g^2_{\rm
YM}$ is small, the entropy of the gas of massless gauge bosons and
their superpartners at temperature $T$ is simply
\beq
S_3 =\frac{2\pi^2}{3} k^2 V T^3 \ ,
\eeq
where $V$ is the spatial volume of the 3-branes.  What happens when
the effective coupling is large?  In this case one can use the type
IIB supergravity solution describing the $k$ coincident 3-branes.
These classical solutions with the asymptotic geometry and quantum
numbers appropriate for $k$ coincident 3-branes contain a
non-extremality parameter upon which their masses and horizon areas
depend.  If we associate the area of the horizon with BH entropy, we
can derive a temperature by taking an appropriate derivative.  By this
procedure, the supergravity picture yields the strong coupling form of
the entropy.  In this case, the entropy agrees with the preceding weak
coupling formula up to a numerical prefactor 3/4, which is then a
prediction for the strong coupling behavior of the theory.

For $k$ coincident D2-branes the UV theory (in the decoupling limit)
is again a weakly-coupled U($k$) SYM theory, so the UV entropy again
scales as $k^2$.  However we are interested in the IR entropy since,
as we will argue in the next subsection, the physically motivated
temperatures are the ambient de~Sitter temperatures which are small
(vanishingly small as $\La_O \to \La_{\rm obs}$).  In the IR the SYM
theory on the (2+1)-dimensional world-volume becomes strongly coupled,
so one must switch over to the supergravity description. As shown in
Ref.~\cite{malda}, the theory flows in the far IR to that of the M2
brane with BH entropy inferred from the horizon area given
by~\cite{KT}
\beq
S_2 \simeq k^{3/2} A T^2  \ ,
\label{Stwobrane}
\eeq
with $A$ the 2-brane area.  This strongly suggests that such
strongly-coupled brane configurations support ${\cal O}(k^{3/2})$
light degrees of freedom, though the physical nature of these degrees
of freedom remains somewhat mysterious.  Thus the probability for a
semi-classical process involving $k$ coincident D2-branes is
multiplied by a density of states factor of the form $N_k \sim
\exp\bigl(k^{3/2} A T^2\bigr)$ in the IR limit $T\to 0$.

The branes of interest to us are effective 2-branes arising from
the wrapping of, say, a D4-brane on a 2-cycle, $a_2$, or in the M
theory picture an M5-brane wrapped on $S^1\times a_2$.  Forgetting
for a moment about the wrapping on a cycle, $k$ coincident D4-branes
have a (4+1)-dimensional SYM theory which now flows to a free theory
in the IR, so the entropy would scale as $k^2 V_4 T^4$ (the M5-branes
have a rather unusual, and microscopically not fully understood,
scaling $S\sim k^3$).  In the case of the wrapped branes, the
excitations in the wrapped directions are massive and not excited at
the low temperatures we consider, so the entropy scales as $S\sim A
T^2$ with $A$ being the area of the 2-brane in our extended
(3+1)-dimensional spacetime.  Such behavior is correct until the
temperature $T$ falls below the inverse linear size $1/b$ of the
nucleated brane, below which point the density of states factor just
counts the number of zero modes.  {}From the preceding discussion, we
expect the scaling of the entropy with $k$ in this case to lie between
$k^2$ and $k^{3/2}$.  A more exact treatment requires an additional
analysis of the way in which the exponent scales as we approach the
IR, as the physically motivated temperature is small but non-zero, and
it is incorrect to scale infinitely far into the IR.  In the case
of a degenerating cycle, however, since the near degeneration implies
that the world-volume theory of the effective 2-brane in (3+1)
dimensions has a bare coupling that is large, we regard the lower
value $\be=3/2$ as being more likely.

In any case, if we accept this reasoning, the probability for a
semi-classical process involving $k$ coincident 2-branes must be
multiplied by an appropriate density of states factor of the form
\beq
N_k \sim \exp\bigl(k^{\be} A T^2\bigr) \ ,
\label{eq:dfactor}
\eeq
for an appropriate temperature $T$, with an exponent $\be$ that likely
lies between $\be=2$ and $\be=3/2$.  (Although we will not utilize it
here, there might also be the possibility of $k^3$ scaling in the
M5-brane limit.)  As we will show below, a larger $\be$ exponent
implies more complete saltation, so we will adopt the more
`conservative' value of $\be=3/2$ as our canonical choice.

\subsection{Temperature: ambiguity, black hole analogy}

The only temperatures intrinsic to our scenario are the de~Sitter
temperatures
\beq
T_{O,I} = \frac{H_{O,I}}{2\pi} =
\frac{1}{2 \pi} \sqrt{\frac{\La_{O,I}}{3}} \ .
\eeq
Ambient ordinary matter might supply a much higher temperature (see
Sec.~\ref{sec:discussion}), but the branes are in very poor thermal
contact with it.  (Ambient D-matter, {\em i.e.}, the light particles
mentioned above, might supply a better coupled temperature bath, but
we shall not pursue this possibility.)

If the initial cosmological term is much larger than the change
brought about by the $k$-bounce,
\beq
\La_O \gg k\rho_2 c_O \ ,
\label{lalarge}
\eeq
then the de~Sitter temperatures before and after nucleation are almost
identical, $T_O \simeq T_I$, and we may use either one in calculating
the density of states factor.

In the case that a given transition produces large changes in the
effective cosmological constant, an ambiguity arises.  One possibility
is that the temperature scale for tunneling from a highly curved (high
temperature) de~Sitter space to a less curved de~Sitter space (or even
to a flat or anti-de~Sitter space) is substantially set by the high
temperature.  In this case one would take
\beq
T \sim T_O
\eeq
in the density of states factor of Eq.~(\ref{eq:dfactor}).  We will
consider the dynamics of this possibility in Sec.~\ref{sec:scenario1}.

However, when the change in the nominal de~Sitter temperature is
comparable to the temperature itself, the thermal description of the
tunneling process is internally inconsistent.  A similar situation has
been encountered before, in black hole physics~\cite{sandip}.  The
problem arises in its most acute form for near-extremal holes, as the
temperature approaches zero.  If one uses the temperature of the
initial hole, one finds a significant probability for radiating a
quantum that will take the hole past extremality to a naked
singularity.  A more refined analysis~\cite{KW,kraus} shows that
radiation is {\it not} thermal with regard to the initial temperature,
and in particular that radiation beyond extremality is forbidden.

If we make an analogy between maximally homogeneous cosmologies and
black holes based on their temperatures, then de~Sitter spaces
correspond to ordinary holes, flat space corresponds to an extremal
hole, and anti-de~Sitter spaces to naked singularities.  This analogy
suggests, in view of the previous paragraph, that we should not
consider finite temperature branes that mediate transitions from
de~Sitter to anti-de~Sitter spaces.  A crude working hypothesis, which
interpolates smoothly to this suggestion, is that in the density of
states factor of Eq.~(\ref{eq:dfactor}), we should, instead of $T\sim
T_O$, employ the geometric mean of the de~Sitter temperatures
\beq
T \sim \sqrt{T_O T_I} \ .
\eeq
The dynamics of this possibility is explored in
Sec.~\ref{sec:scenario2}.

\subsection{A catastrophe for $T\sim T_O$?}

If the analogy to black hole results holds, tunneling to (and through)
anti-de~Sitter space is excluded.  On the other hand, if the
temperature for transitions is effectively set by $T_O$, multi-bounce
transitions with arbitrarily large $k$ are possible and must be
considered.

The action for $k$-bounce tunneling is given by
Eqs.~(\ref{bounce})--(\ref{relations}), with $\rho_2 \to k\rho_2$ and
$\tau_2 \to k\tau_2$.  The action is monotonically increasing as $k$
increases, and has the limiting behavior
\beq
B = \frac{24\pi^2 M^2}{\La_O}\ , \quad k \to \infty \ .
\eeq
The limit $k\to \infty$ therefore appears problematic, since $P \sim
N_k e^{-B}$ approaches a constant for large $k$.  Note, however, that
as $k \to \infty$, the bubble size $b \sim \frac{1}{k} \to 0$.  Thus,
the apparent `instability' is toward creation of highly curved branes.
We do not expect the action of Eq.~(\ref{Sm}) to be valid in the
regime where the brane curvature far exceeds the brane
tension.\footnote{We thank R.~Sundrum for emphasizing this point to
us.}  Presumably, a calculation of the tunneling probability with the
degrees of freedom appropriate for this regime will be well-behaved.

\section{Scenarios}
\label{sec:scenarios}

Let us now gather the pieces and attempt to envisage how --- and
whether! --- they may be assembled into a complete scenario.

The cosmological constant evolves from some initial value through
multi-bounce transitions.  The probability for such transitions is
\beq
P \sim e^D e^{-B} \ .
\eeq
The bounce action $B$ is given by
Eqs.~(\ref{bounce})--(\ref{relations}), with $\rho_2 \to k\rho_2$ and
$\tau_2 \to k\tau_2$, where $k$ is the bounce number.  The density of
states prefactor is specified by $D=k^{\be} A T^2$, where $A=4 \pi
b^2$ is the 2-brane area, and $T$ is the temperature.  For
concreteness, we will assume the low `conservative' value of $\be =
3/2$, but the qualitative features of the following analysis hold more
generally, for example, for $\be=2$ or larger.

As noted above, the temperature $T$ is not under good theoretical
control. We will therefore explore both of the broad alternatives
mentioned previously.

\subsection{$T \sim T_O$, 1-step relaxation}
\label{sec:scenario1}

We first consider the possibility that the temperature is given by the
scale of the initial (outside) de~Sitter temperature, so $T \sim T_O
\sim \sqrt{\La_O}$.  The tunneling probability from a given background
configuration is then fixed in terms of the initial effective
cosmological constant $\La_O$, the initial field strength $c_O$, and
the 2-brane charge density and tension, parameterized by $\rho_2$ and
$x \equiv \tau_2 / \rho_2$.

We begin with some bare cosmological constant $\la<0$.  Assume that
the initial field strength gives a similar contribution to the
effective cosmological constant, so $\La_O, c_O^2 \sim |\la|$.  We
assume also a very small charge density $\rho_2 \sim \La_{\rm obs} /
\sqrt{|\la|}$, consistent with the naturalness condition discussed in
Sec.~\ref{sec:naturalness}, and $x\sim 1$.

With such initial conditions and brane properties, the maximal bounce
action is $B \sim 1/|\la|$, while the degeneracy factor may be as
large as $D \sim \la^2 / \La_{\rm obs}^2$.  Recall that $\La_{\rm obs}
\sim 10^{-120}$, while $\la$ is plausibly in the range of 1 (for
Planck scale cosmological constants) to $10^{-60}$ (for low-scale
supersymmetry breaking at the TeV scale).  Thus, the degeneracy
enhancement overpowers the bounce action suppression, and tunneling
proceeds rapidly.

It is not difficult to show that $D$ is maximized for $k \rho_2 \sim
c_O$, {\em i.e.}, for field strength step sizes of the right order to
neutralize the field strength contribution to the effective
cosmological constant.  Indeed the most probable tunneling events
nucleate bubbles of deep anti-de~Sitter space.  Such events produce
small, short-lived interior universes, so the meaning of `probable' in
this context must be carefully qualified.  Among universes that live a
long time and even remotely resemble ours, the exponentially most
favored are those closest to having zero effective cosmological term.

This scenario invokes a form of the anthropic principle.  It is a
uniquely weak one, however, in the following sense.  Anthropic bounds
on the cosmological term are highly asymmetrical~\cite{weinberg}.  For
positive cosmological terms, the formation of sufficiently large
gravitational condensations requires cosmological terms below $\sim
100$ in units of $\rho_c$, the critical density.  For negative
cosmological terms, the lifetime of the universe requires cosmological
terms roughly above $-1$.  Thus if the spacing between allowed
near-zero saltatory values of the cosmological term in units of
$\rho_c$ is, say, 3 and allows the values $\ldots, -2, 1, 4, \ldots$,
then among values that can be experienced by sentient observers, 1 is
by far the most likely.

Now suppose such a transition to nearly flat space has occurred.  As
discussed in Sec.~\ref{sec:stability}, absolute stability is possible
only for anti-de~Sitter spaces, and only for extremely large $x$.
However, absolute stability is not required on empirical grounds.  We
need only require that the effective value of the cosmological term is
at present stable on cosmological time scales.  For a starting
effective cosmological constant of $\La_O \sim \La_{\rm obs}$, the
degeneracy factor is highly suppressed by small $T$, and the bounce
action is dominant.  The stability of the vacuum is then determined
solely by $B$.

Since the bounce action can become truly infinite for slightly
anti-de~Sitter spaces, we anticipate, by continuity, that it can
become very large even for slightly de~Sitter spaces.  The least
suppressed transition (and so most dangerous from the point of view of
vacuum instability) is that mediated by $k=1$.  For small $\rho_2$ and
small $\La_O$, the $k=1$ bounce action is
\beq
B \approx \frac{27\pi^2 x^4}{2} \left[\frac{\rho_2}{c_O^3}
- \frac{\rho_2^2}{c_O^2 \La_O } \right] \ , \quad
\rho_2, \sqrt{\La_O} \ll c_O \ .
\label{zaza}
\eeq
Neglecting numerical factors, we find
\beq
B \sim  x^4 \La_{\rm obs} / \la^2 \ .
\eeq
If we now take $|\la| \sim 1$, then the action is very small, and
there is no effective stability.

On the other hand, if supersymmetry is broken at a low scale, then we
expect $|\la| \ll 1$.  Let us inquire when $B \gsim 1$.  This
translates into
\beq
|\la_{\rm halting}| \lsim x^2 \sqrt {\La_{\rm obs}} \ ,
\label{halting}
\eeq
or, restoring the mass units,
\beq
|\la_{\rm halting}| M^2 \lsim
x^2 (2 \times 10^{-3} \ev)^2 (2.4 \times 10^{18} \gev)^2
\sim x^2\, (2 \tev)^4 \ .
\eeq
In models where supersymmetry breaking is transmitted with little
suppression to the Standard Model fields (or is even present in the
observable sector itself), it is reasonable to expect the
supersymmetry breaking scale to set the scale of the bare cosmological
constant, so
\beq
M_{\rm weak}^4 \sim M_{\rm SUSY}^4
\sim |\la_{\rm halting}| M^2 \ .
\eeq
Given the present experimental lower bounds on the supersymmetry
breaking scale, we find then that the stability of the vacuum in this
scenario requires low scale supersymmetry breaking, and relates the
cosmological constant, Planck, and weak scales according to
\beq
M_{\rm weak}^2 \sim (10^{-3} \ev) (M_{\rm Planck}) \ ,
\eeq
in accord with observation.

In a more careful analysis, one may require $B \gg 1$ for stability.
However, the required supersymmetry breaking scale will not differ
significantly from the above estimates, as $B$ goes as the inverse 8th
power of the energy scale appearing in $|\la| M^2$.  Finally, note
that we have assumed the qualitative validity of the bounce action
Eq.~(\ref{bounce})--(\ref{relations}) throughout this section.  For
large $\lambda \gg 10^{-60}$, this may not be appropriate, as the
brane curvature may be much greater than the brane tension.  However,
we have checked that in the case of greatest interest to us with
$\lambda \sim 10^{-60}$, the brane curvature never greatly exceeds the
brane tension, and the analysis above is under reasonable control.

\subsection{$T \sim (T_I T_O)^{1/2}$, multi-step relaxation}
\label{sec:scenario2}

Motivated by the black hole analogy, we now consider an effective
temperature that is the geometric mean of the initial and final
de~Sitter temperatures.  In this case, tunneling to (and through)
anti-de~Sitter space is forbidden by fiat.  However, the requirements
of rapid tunneling to the observed cosmological constant and its
stability are non-trivial constraints, and we now investigate their
implications.

As in the previous scenario, we consider initial conditions $c_O^2,
\La_O \sim |\la|$.  Now, however, the density of states factor $D$ is
typically maximized for $\La_I$ within an order of magnitude of
$\La_O$.  To see this, a very rough estimate may be obtained by
neglecting the bubble radius dependence on $k$ and approximating
$\La_O - \La_I = (2 \rho_2 c_O - \rho_2^2)/2 \sim \rho_2 c_O$.  We
then have $D \propto \sqrt{\La_O \La_I} (\La_O-\La_I)^{3/2}$, which is
maximized for $\La_I = \frac{1}{4} \La_O$.

For $\La_I \sim \La_O$,
\beq
D_{\rm max} \sim \frac{\La_O^{5/2}}{|\la|^{7/4}\rho_2^{3/2}}
\ .
\eeq
It is not hard to verify that this degeneracy factor dominates the
bounce suppression when $\La_O \sim |\la|$.  Thus, initially the
effective cosmological constant tunnels rapidly as in the previous
scenario, but in contrast to the previous case, the cosmological
constant relaxes through several steps, with values roughly following
a geometric series.

The effective cosmological constant will relax as described until
$\La_O \ll c_O^2$, when Eq.~(\ref{zaza}) holds.  At this point,
the condition that tunneling continue is the requirement $D_{\rm max}
\gsim B$, or, since $B \sim \La_O / \la^2$,
\beq
\La_O^{3/2} \gsim |\la|^{-1/4} \rho_2^{3/2} \ .
\label{dmax}
\eeq
For vanishing $\rho_2$, tunneling may continue to arbitrarily small
$\La_O$.  However, if we require stability from $B \gsim 1$, we
find, from Eq.~(\ref{zaza}),
\beq
\rho_2 \gsim c_O^3 \sim |\la|^{3/2} \ ,
\label{rho0}
\eeq
so $\rho_2$ cannot be arbitrarily small.  Combining Eqs.~(\ref{dmax})
and (\ref{rho0}), we find that tunneling stops when
\beq
\La_O \gsim |\la|^{4/3} \ .
\eeq
Thus, even for $|\la| \sim 10^{-60}$, although the effective
cosmological constant is reduced by a factor of $10^{20}$, one
membrane cannot suppress it to the observed value.

In general, however, it is important to note that several different
2-branes with various fundamental charge densities may be expected to
arise.  Suppose that another brane begins nucleating as the first
membrane reaches its endpoint.  The initial conditions for this new
membrane are identical to those for the first brane, except that now
the role of the initial bare cosmological constant is played by $\La_O
\sim |\la|^{4/3}$. For appropriate charge densities, $n$ branes may
reduce the cosmological constant to $|\la|^{(4/3)^n}$.  For $|\la|
\sim 10^{-60}$, three branes are sufficient to reduce the cosmological
constant to its observed value.

So far we have considered only the `conservative' $\be=3/2$ case.  For
larger values of $\be$ more complete relaxations of the cosmological
term are possible.  For general $\be$, a single membrane may relax the
cosmological constant to $\Lambda_O \gsim |\lambda|^{2 - \be^{-1}}$.
Thus, even for the $\be =2$ case, only two stages are required.  Note
also that in these multi-brane scenarios, in principle quite complex
dynamics can arise, with periods of slow relaxation interspersed with
more rapid changes.

\section{Summary and Discussion}
\label{sec:discussion}

On very general grounds, it is appealing to think that relaxation of
the cosmological term might be associated with very special degrees of
freedom that have no conventional couplings to matter, and no
conventional kinetic energy, but respond only to 3+1 dimensionally
uniform form fields and, of course, to gravity.  Several difficulties
in such an approach must be addressed. The phenomenologically required
energy scale is very small and not easily manufactured out of
conventional energy scales.  In addition, in any reasonable scenario,
the cosmological constant must relax sufficiently quickly from high
scales, but must be stable on cosmological time scales at its present
value.

String theory provides a promising microscopic framework for such a
mechanism. The necessary degrees of freedom are naturally supplied by
string/M theory branes, and the dependence of brane properties on
compactification appears, in principle, to be capable of producing a
very small scale.  We have also identified a candidate mechanism, the
enhancement of multi-step jumps due to large density of states
factors, which typically leads to large tunneling probabilities.
Finally, as we have seen, the absolute stability of certain
anti-de~Sitter universes implies that near flat universes may be also
be sufficiently stable.

We have considered two representative scenarios differing in the
treatment of the effective temperature entering the density of states
factor.  In the simplest scenario, with $T \sim T_O$, the
exponentially most probable transition, excluding extremely
short-lived universes, is to universes that are most nearly flat.  By
requiring that this new vacuum be sufficiently stable, we derived a
non-trivial constraint for a mechanism of this kind.  This
constraint provides an intriguing relationship, between the
supersymmetry breaking scale and the geometric mean of the present-day
effective cosmological constant and Planck scales:
\begin{equation}
M_{\rm SUSY}^2 \lsim (10^{-3}\ev) (M_{\rm Planck}) \ .
\end{equation}
Large supersymmetry breaking scales are thereby excluded, and the
largest possible scale is plausibly though not necessarily satisfied
in Nature.

In reaching this relation, we have assumed the bare cosmological
constant to be of order the supersymmetry breaking scale.  Indeed,
while the bare cosmological term appears as a free parameter in
supergravity, in string/M theory, phenomenologically interesting
models with unbroken supersymmetry have zero cosmological term.  In
this context, it is therefore reasonable to expect that the relevant
scale for the bare cosmological term is indeed the supersymmetry
breaking scale.  Many models of supersymmetry breaking invoke `hidden
sectors' with a characteristic mass scale much larger than the TeV
scale.  Unless the hidden sector contribution to the bare cosmological
term is somehow suppressed to this TeV scale, in our simplest scenario
the vacuum will be unstable.

Alternatively, an analogy with black holes suggests a richer dynamics,
in which flat space plays a distinguished role and tunneling to
anti-de~Sitter space is forbidden.  Within this circle of ideas, and
in contrast to the previous scenario, we found that the cosmological
constant relaxes through a several jumps, roughly following a
geometric series.  The constraint of stability limits the range over
which the cosmological constant may be relaxed by any given membrane.
However, two or more types of branes with radically different scales
may relax the cosmological constant to within observational bounds,
and appeal to the anthropic principle may be avoided.

Our work so far is very seriously incomplete, in that we have not
attempted to incorporate it into a realistic ({\em e.g.},
Friedmann-Robertson-Walker) cosmological model including matter.
Thus, in particular, we have not addressed the dynamics of relaxation
following a phase transition.  There is a potential problem here,
since, if after relaxation to zero effective cosmological term a later
matter phase transition drives it negative, recovery may be difficult.
We note also that the existence of very light membrane degrees of
freedom may have a variety of observational and experimental
consequences.  We reserve discussion of these issues for a future
publication.

We also require, for our dynamics, compactification schemes that
produce the desired brane properties.  Most model building in string/M
theory has been based, implicitly or explicitly, on the paradigm of
minimizing a potential.  This has always been problematic within a
quantum theory of gravity, but it was not clear what could replace it.
The saltatory mechanism suggests a different principle, based on the
dynamics of relaxation of the cosmological term.  Whether this
principle, or any other, is powerful enough to select uniquely a
vacuum as complex as the one we observe remains to be seen.

\section{Acknowledgments}

We are very grateful to Luis Alvarez-Gaume, Paul Aspinwall, Raphael
Bousso, Amit Giveon, Alex Kusenko, Juan Maldacena, Emil Martinec,
Peter Mayr, Greg Moore, Yaron Oz, Joe Polchinski, Raman Sundrum, and
Claudio Teitelboim for discussions.  The work of JLF is supported in
part by the Department of Energy under contract DE--FG02--90ER40542
and through the generosity of Frank and Peggy Taplin.  JMR wishes to
thank the Alfred P. Sloan Foundation for the award of a Fellowship,
and the US Department of Energy for an Outstanding Junior Investigator
Award.  The work of SS is supported in part by the William Keck
Foundation and by NSF grant No. PHY--9513835.  The work of FW is
supported in part by the Department of Energy under contract
DE--FG02--90ER40542 and by the National Science Foundation under grant
PHY--9513835.

\section*{Note}

As this manuscript was being completed, we learned of independent work
by Bousso and Polchinski~\cite{BP} proposing quite a different
scenario for fixing the cosmological term in string theory through a
generalization of the Brown-Teitelboim mechanism.  They do not utilize
degenerating cycles nor enhanced density of states factors, and
instead invoke the anthropic principle in an essential way.  We thank
them for conversations regarding our respective approaches.

%
\def\NPB#1#2#3{Nucl. Phys. {\bf B#1} #2 (#3)}
\def\PLB#1#2#3{Phys. Lett. {\bf B#1} #2 (#3)}
\def\PLBold#1#2#3{Phys. Lett. {\bf#1B} #2 (#3)}
\def\PRD#1#2#3{Phys. Rev. {\bf D#1} #2 (#3)}
\def\PRL#1#2#3{Phys. Rev. Lett. {\bf#1} #2 (#3)}
\def\PRT#1#2#3{Phys. Rep. {\bf#1} #2 (#3)}
\def\ARAA#1#2#3{Ann. Rev. Astron. Astrophys. {\bf#1} #2 (#3)}
\def\ARNP#1#2#3{Ann. Rev. Nucl. Part. Sci. {\bf#1} #2 (#3)}
\def\MPL#1#2#3{Mod. Phys. Lett. {\bf #1} #2 (#3)}
\def\ZPC#1#2#3{Zeit. f\"ur Physik {\bf C#1} #2 (#3)}
\def\APJ#1#2#3{Ap. J. {\bf #1} #2 (#3)}
\def\AP#1#2#3{{Ann. Phys. } {\bf #1} #2 (#3)}
\def\RMP#1#2#3{{Rev. Mod. Phys. } {\bf #1} #2 (#3)}
\def\CMP#1#2#3{{Comm. Math. Phys. } {\bf #1} #2 (#3)}
\relax
%
\newcommand{\journal}[4]{{ #1} {\bf #2}, #3 (#4)}
\newcommand{\hepth}[1]{{hep-th/#1}}
\newcommand{\hepph}[1]{{hep-ph/#1}}
\newcommand{\grqc}[1]{{gr-qc/#1}}
\newcommand{\astro}[1]{{astro-ph/#1}}
%


\begin{thebibliography}{9}

\bibitem{weinberg}
S. Weinberg, \RMP{61}{1}{1989}.

\bibitem{CC}
S. Coleman, \PRD{15}{2929}{1977}, {\em Erratum},
\PRD{16}{1248}{1977};\\
C. Callan and S. Coleman, \PRD{16}{1762}{1977}.

\bibitem{CdL}
S. Coleman and F. de Luccia, \PRD{21}{3305}{1980}.

\bibitem{abbott}
L. Abbott, \PLB{150}{427}{1985}.

\bibitem{BT}
J. Brown and C. Teitelboim, \PLB{195}{177}{1987};\\
\NPB{297}{787}{1988}.

\bibitem{others}
R. Bousso and A. Chamblin, \PRD{59}{063504}{1999}, \hepth{9805167};\\
M. S. Bremer \etal, \NPB{543}{321}{1999}, hep-th/9807051.

\bibitem{HT}
See, {\em e.g.}, S. Hawking and N. Turok, \PLB{432}{271}{1998},
\hepth{9803156}.

\bibitem{reviews}
J.~D.~Cohn,
astro-ph/9807128;
S. Carroll, astro-ph/0004075.

\bibitem{townsend}
P. K. Townsend, \PLB{350}{184}{1995}, hep-th/9501068;\\
E. Witten, \NPB{443}{85}{1995}, hep-th/9503124.

\bibitem{Mlectures}
For excellent introductory lectures see:\\
P. K. Townsend, hep-th/9612121;\\
M. J. Duff, hep-th/9611203.

\bibitem{JPbook}
J. Polchinski, {\em String Theory: Vols. I and II}, (Cambridge
University Press, Cambridge, 1998).

\bibitem{vafa}
C.~Vafa, \NPB{469}{403}{1996}, \hepth{9602022}.

\bibitem{sethione}
S. Sethi, C. Vafa, and E. Witten, \NPB{480}{213}{1996},
\hepth{9606122}.

\bibitem{beckers}
K. Becker and M. Becker, \NPB{477}{155}{1996}, \hepth{9605053}.

\bibitem{sethitwo}
K. Dasgupta, G. Rajesh, and S. Sethi, JHEP {\bf 9908} 023 (1999),
\hepth{9908088}.

\bibitem{greene}
B. R. Greene, K. Schalm, and G. Shiu, \hepth{0004103}.

\bibitem{polRR}
J. Polchinski, \PRL{75}{4724}{1995}, \hepth{9510017}.

\bibitem{Dbranes}
For excellent introductions see:\\
J. Polchinski, \hepth{9611050};\\
C. Bachas, \hepth{9806199}.

\bibitem{candelas}
P. Candelas and D. Raine, \NPB{248}{415}{1984}.

\bibitem{PS}
J. Polchinski and A. Strominger, \PLB{388}{736}{1996},
\hepth{9510227}.

\bibitem{michelson}
J. Michelson, \NPB{495}{127}{1997}, \hepth{9610151}.

\bibitem{taylorvafa}
T. Taylor and C. Vafa, \PLB{474}{130}{2000}, \hepth{9912152}.

\bibitem{mayr00}
P. Mayr, \hepth{0003198}.

\bibitem{Mquant}
J. Schwarz, \PLB{367}{97}{103}, \hepth{9510086};\\
S. de~Alwis, \PLB{388}{291}{1996}, \hepth{9607011}.

\bibitem{AAHDD}
I. Antoniadis \etal, \PLB{436}{257}{1998}, \hepph{9804398}.

\bibitem{gauge}
I. Antoniadis and C. Bachas, \PLB{450}{83}{1999}, \hepth{9812093};\\
N. Arkani-Hamed \etal, \hepth{9908146};\\
K. Dienes, E. Dudas, and T. Gherghetta, \NPB{537}{47}{1999},
\hepph{9806292}.

\bibitem{conifold}
A. Strominger, \NPB{451}{96}{1995}, \hepth{9504090}.

\bibitem{concandelas}
P.~Candelas, P.~S.~Green and T.~Hubsch,
\NPB{330}{49}{1990}.

\bibitem{hayakawa}
Y. Hayakawa, 
alg-geom/9507016.

\bibitem{morrison}
P.~Candelas, D.~Diaconescu, B.~Florea, D.~R.~Morrison and G.~Rajesh,
hep-th/0009228.

\bibitem{mayr96}
P. Mayr, \NPB{494}{489}{1997}, \hepth{9610162}.

\bibitem{mayrprivate}
P. Mayr, private communication.

\bibitem{joke}
F. Wilczek, \journal{Phys. Scripta}{T36}{281}{1991}.

\bibitem{BK}
C. Bachas and E. Kiritsis, \journal{Nucl. Phys. Proc.
Suppl.}{B55}{194}{1997}, \hepth{9611205};\\
C. Bachas \etal, \NPB{509}{33}{1998}, \hepth{9707126}.

\bibitem{kiritsis}
For excellent lectures, see:\\
E. Kiritsis, \hepth{9906018}.

\bibitem{KT}
I. Klebanov and A. Tseytlin, \NPB{475}{164}{1996}, \hepth{9604089}.

\bibitem{malda}
N. Itzhaki \etal, \PRD{58}{046004}{1998}, hep-th/9802042.

\bibitem{sandip}
J. Preskill \etal, \journal{Mod. Phys. Lett.}{A6}{2353}{1991}.

\bibitem{KW}
P. Kraus and F. Wilczek, \NPB{433}{403}{1995}, \grqc{9408003}; \\
\NPB{437}{231}{1995}, \hepth{9411219}.

\bibitem{kraus}
E. Keski-Vakkuri and P. Kraus, \NPB{491}{249}{1997}, \hepth{9610045}.

\bibitem{BP}
R. Bousso and J. Polchinski, \hepth{0004134}.


\end{thebibliography}
\end{document}